\documentclass [aps,prl,twocolumn,longbibliography,superscriptaddress]{revtex4-2}

\usepackage{graphicx}
\usepackage{subfigure}
\usepackage{amsmath}
\usepackage{amsthm}
\usepackage{amssymb}
\usepackage{hyperref}
\usepackage{xcolor}
\usepackage{textcomp}
\usepackage[normalem]{ulem}
\usepackage{orcidlink}
\hypersetup{colorlinks=true, linkcolor=black, citecolor=black, urlcolor=blue}

\def\be{
\begin{equation}}
  \def\ee{
\end{equation}}

\begin{document}

%=========================================================================
\title{Entanglement Entropy of Free Fermions on Random Fractal Lattices}

\author{Arkadiusz Kosior\,\orcidlink{0000-0002-5039-1789}}
\affiliation{Institut f\"ur Theoretische Physik, Universit{\"a}t Innsbruck, A-6020~Innsbruck, Austria}
\affiliation{Institute of Physics, Maria Curie-Skłodowska University, 20-031 Lublin, Poland}
\author{Jia Wang\,\orcidlink{0000-0002-9064-5245}}
\affiliation{Centre for Quantum Technology Theory, Swinburne University of Technology, Melbourne 3122, Australia}

\begin{abstract}
  Random fractal lattices provide a geometrically disordered setting in which quantum correlations can be shaped by noninteger dimensionality rather than onsite randomness. We investigate the entanglement properties of noninteracting fermions on random fractal lattices generated by a stochastic growth algorithm. By varying the growth parameter and adding missing links with probability $p$, we tune the Hausdorff and spectral dimensions while keeping the system free of onsite disorder. For ground states at different fillings, we compute the bipartite entanglement entropy of subregions defined by graph distance and analyze its scaling with subsystem size. Over a broad parameter range, we find robust power-law behavior governed primarily by the Hausdorff dimension, consistent with a generalized area law and without the logarithmic enhancement familiar from Euclidean free fermions. We also study entanglement growth following a global quench from an uncorrelated checkerboard state and uncover an asymptotic scaling collapse in which the subsystem-size dependence is governed by the Hausdorff dimension, while the temporal evolution is governed by the spectral dimension. The resulting dynamics are logarithmically slow over an extended intermediate-time window. These results show that geometric randomness alone can generate both nontrivial ground-state entanglement structure and slow quantum-information spreading in free-fermion systems.
\end{abstract}

\maketitle

\textit{Introduction.---} Quantum entanglement is a form of non-local correlation with no classical equivalent \cite{Horodecki2009,bengtsson_zyczkowski_2017}. It represents the fundamental inseparability of individual components within a many-body system, yielding a collective state where constituent parts cannot be described independently. Beyond its theoretical interest, entanglement serves as the foundational resource driving quantum-enhanced technologies, including quantum teleportation \cite{Bennett1993} and high-precision quantum metrology~\cite{Leibfried2004,Colombo2022}. Consequently, intense effort is focused on the scalable generation of interaction-induced entanglement. Traditionally engineered via coherent unitary dynamics like one-axis twisting \cite{Kitagawa1993}, a recent finding demonstrates that this form of entanglement can also be generated by tailoring dissipative quantum trajectories \cite{Hotter2025}. Conversely, many-body ground-state entanglement emerges naturally from the collective, equilibrium configurations of a many-body Hamiltonian rather than active dynamical manipulation. This intrinsic entanglement can be quantified by the entanglement entropy (EE), defined as the von Neumann entropy of the reduced density matrix for a given spatial subregion. Beyond merely quantifying bipartite correlations, EE offers a compact probe of universal low-energy structures that often remain hidden from conventional local observables \cite{Plenio2010RMP,Laflorencie2016PhysRep}. Consequently, it has become a standard diagnostic tool for characterizing critical behavior \cite{Vidal2003PRL,Vedral2008RMP}, diagnosing topological order \cite{Kitaev2006PRL,Levin2006PRL}, and identifying the effective dimensionality of low-energy excitations \cite{Swingle2010PRL}.

For a spatial subregion of linear size $L$, EE typically obeys an area law in the ground state of a gapped local Hamiltonian, scaling as $S \sim L^{d-1}$ in $d$ spatial dimensions~\cite{Plenio2010RMP, Laflorencie2016PhysRep}. Gapless systems, however, can violate this scaling. A paradigmatic example is free fermions on translationally invariant Euclidean lattices, for which EE acquires a logarithmic enhancement, $S \sim L^{d-1}\log L$~\cite{Klich2006PRL,WolfPRL2006, Hastings2007JSM, Schollwock2006PRA, Swingle2010PRL}. This form was proposed by Gioev and Klich using the Widom conjecture~\cite{Klich2006PRL}, later established rigorously for free Fermi gases and R\'enyi entropies~\cite{Leschke2014PRL}, and given a physical interpretation via multidimensional bosonization, where the Fermi surface acts as a continuum of effectively one-dimensional gapless modes that contribute additively to the entanglement~\cite{Yang2012PRX}.  
Extending these results beyond integer-dimensional Euclidean systems has attracted growing interest. Already in Ref.~\cite{Klich2006PRL}, it was noted that irregular or fractal boundaries can modify the scaling behavior, leading to bounds of the form $S \gtrsim L^{d-\beta}$, where $\beta \in (0,1)$ characterizes boundary regularity. Reduced connectivity can also qualitatively alter entanglement scaling; for example, free fermions coupled through point contacts exhibit $S \sim m \log L$, where $m$ is the number of contacts \cite{Levine2014PRA}. More recently, entanglement properties of fermions on fractal lattices have been explored. For gapless fermions embedded in Euclidean lattices, a Widom-type logarithmic enhancement may persist, reflecting the presence of extended modes inherited from the embedding geometry. In contrast, for gapped systems defined intrinsically on fractal lattices, a generalized area law $S \sim L^{d_{\mathrm{b}}}$ has been observed, where $d_{\mathrm{b}}$ is the Hausdorff dimension of the boundary \cite{Ye2024PRR, Ye2025PRR}.

Another key aspect of entanglement is its dynamical evolution following a quantum quench. In clean systems, entanglement typically grows linearly in time, $S(t)\sim t$, reflecting ballistic quasiparticle propagation \cite{Cardy2005JSM, Calabrese2024PRB, Bertini2024PRB, nonlinear_growth}. In noninteracting Anderson-localized systems, post-quench entanglement growth is strongly suppressed and the entropy saturates quickly to a bounded value set by the localization length, whereas in many-body localized phases it exhibits slow, logarithmic growth, $S(t)\sim \log t$, due to interaction-induced dephasing \cite{Moore2012PRL, Abanin2013PRL, DasSarma2017PRB, Wang2018NJP, Sirker2019PRB}. Despite extensive studies in Euclidean systems, the dynamics of entanglement in systems with fractal geometry remain largely unexplored.

In this work, we study free fermions on random fractal graphs generated by a dielectric-breakdown growth process and investigate both ground-state entanglement and entanglement generation after a quench \cite{Niemeyer1984_dbb,Kosior2017_localization}. These random fractal lattices (RFLs) are statistically self-similar graphs whose typical realizations are characterized by well-defined Hausdorff and spectral dimensions. We find that, although we restrict to gapless ground states, the EE follows an area law determined by the Hausdorff dimension, without the logarithmic enhancement characteristic of Euclidean free fermions. We attribute the absence of the logarithmic correction to the lack of a well-defined extended Fermi surface on a fractal lattice without translational invariance. Following a quench, by contrast, the entanglement entropy exhibits robust logarithmic-in-time growth. Our results show that slow entanglement dynamics can arise purely from geometric hierarchy, providing a mechanism distinct from both Fermi-surface physics and disorder-induced localization.

\begin{figure}[t!]
  \centering
  \includegraphics[width= 0.49\textwidth]{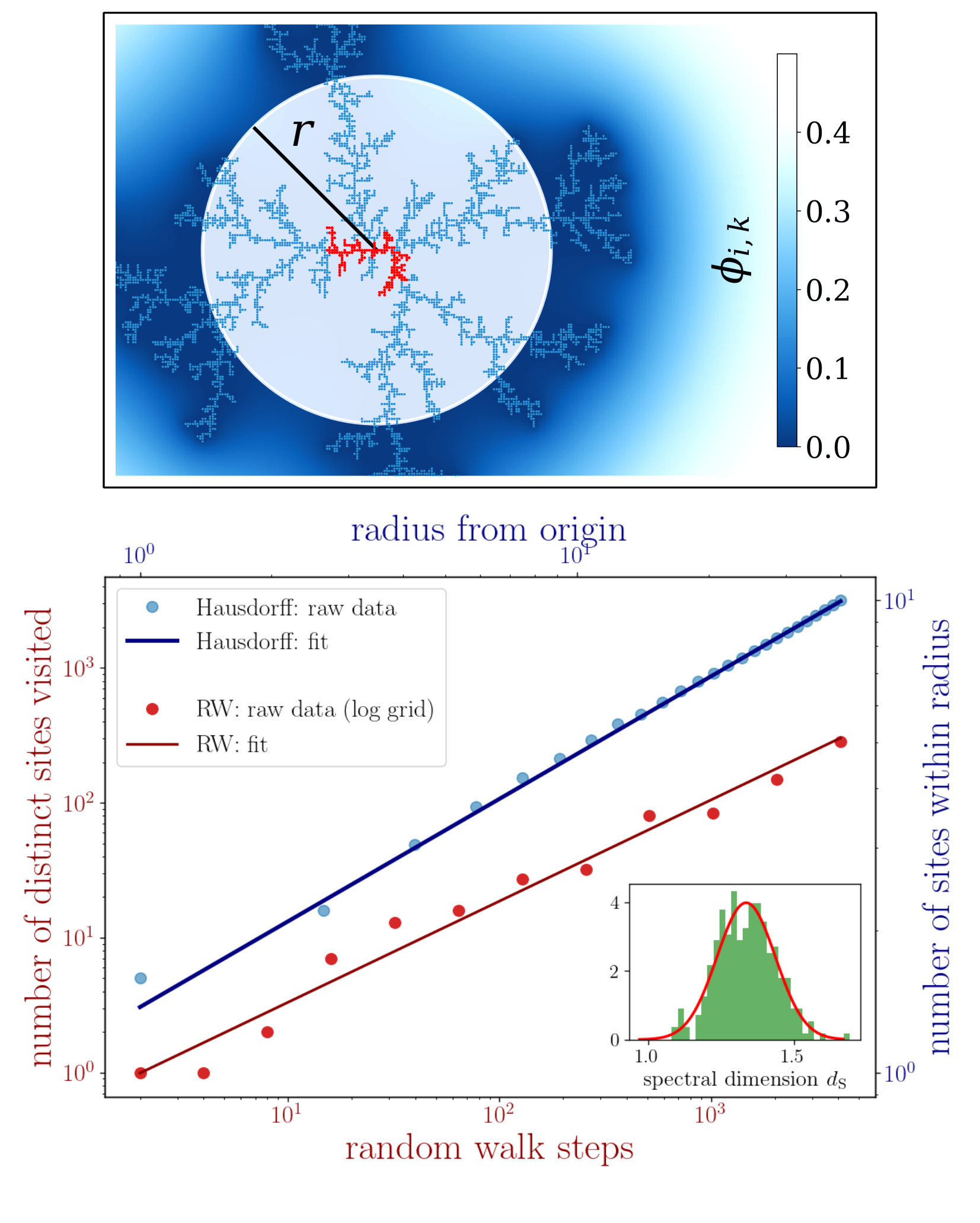}
  \caption{
    (top panel) Example random fractal lattice generated by the growth algorithm. The shaded region indicates the ball of graph radius $r$ around the origin; counting the number of enclosed sites gives the volume $V(r)$ used to extract the Hausdorff dimension $d_{\mathrm H}$. The red sites mark distinct sites $\Omega$ visited during a random walk, which is used to estimate the spectral dimension.
    (bottom panel) Example log--log fits used to extract fractal dimensions from a single lattice realization. Blue and red points show $V(r)\sim r^{d_{\mathrm H}}$ and $\Omega(t)\sim t^{d_{\mathrm s}/2}$, respectively. To reduce finite-size effects and stochastic fluctuations, the walk is run up to $2^{12}$ steps and repeated about 500 times; fitted slopes are then collected into a histogram to estimate $d_{\mathrm s}$.
  }
  \label{fig1}
\end{figure}

\textit{Fractal Growth and the Tight-Binding Model.---} 
In this work we investigate a class of random fractal lattices (RFLs) derived from the dielectric breakdown model~\cite{Niemeyer1984_dbb,Kosior2017_localization}, in which a dielectric discharge pattern forms a fractal structure governed by a single parameter $\eta$ (defined below). More generally, RFLs are random graphs characterized by a self-similar structure in a statistical sense. While individual realizations of these networks differ microscopically, typical samples share a common, universal fractal dimensions: the Hausdorff dimension $d_{\mathrm H}$ and the spectral dimension $d_{\mathrm s}$ \cite{Mandelbrot1967,Rammal1983}.

RFLs that we consider are generated through an iterative probabilistic growth procedure. Starting from an arbitrary seed, new sites are added to the existing cluster from a set of candidate boundary sites $\partial A$. The probability of connecting a specific boundary site $j\in \partial A$ to a site $i\in A$ reads
\begin{equation}
P(i \to j) = \frac{(\phi_j)^\eta}{\sum_{k \in \partial A} (\phi_k)^\eta},
\end{equation}
where the local potential $\phi$, defined on the Euclidean grid, is obtained by self-consistently solving the discrete Laplace equation
\begin{equation}
\phi_{i} = \frac{1}{n} \sum_\delta \phi_{i+\delta}
% \left( \phi_{j+1,k} + \phi_{j-1,k} + \phi_{j,k+1} + \phi_{j,k-1} \right)
\end{equation}
until convergence is reached. Here, the sum runs over nearest-neighbor sites, with $n=4$ ($n=6$) for two- (three-) dimensional grids. Once site $j$ is added, its potential is set to zero, $\phi_{j}=0$, and the Laplace field is recalculated. This process yields a single connected graph without loops, meaning that nearest-neighboring sites on the underlying grid are not necessarily connected.

While the Hausdorff dimension $d_{\mathrm H}$ is a box-counting dimension that characterizes the geometric mass of a graph, the spectral dimension $d_{\mathrm s}$ captures its underlying diffusion properties and effective connectivity. In order to tune the graph connectivity, we add ``missing'' nearest-neighbor links with probability $p$, thereby interpolating between the minimally connected graph at $p=0$, and the fully connected nearest-neighbor graph on the underlying embedding at $p=1$. In this way, the spectral dimension can be varied, to a significant extent, independently of the Hausdorff dimension.

In the following we focus on fractal lattices embedded in two-dimensional Euclidean planes. To characterize the resulting geometries, we numerically estimate both $d_{\mathrm H}$ and $d_{\mathrm s}$. The Hausdorff dimension is extracted from the scaling of the volume $V(r) \sim r^{d_{\mathrm H}}$, where $V(r)$ is defined as the number of sites within graph distance $r$ from a reference point, as illustrated in Fig.~\ref{fig1}.
The spectral dimension $d_{\mathrm s}$, which governs diffusion and connectivity, is determined from unbiased classical random walks, also shown in Fig.~\ref{fig1}. We monitor the mean number of distinct sites visited, $\Omega(t)$, as a function of time $t$, which follows the scaling law
$\Omega(t) \sim t^{d_{\mathrm s}/2}$.
To improve statistical accuracy and reduce finite-size effects, we perform approximately 500 independent walks for each realization and compile histograms of the fitted slopes to determine the distribution of $d_{\mathrm s}$, as shown in Fig.~\ref{fig1}. The extracted mean fractal dimensions for ensembles with fixed $\eta$ and $p$ are summarized in Fig.~\ref{fig2} in the End Matter (EM). In our numerical simulations, we stop the growth algorithm after $N = 5000$ steps.

On this geometric foundation, we study noninteracting spinless fermions described by the tight-binding Hamiltonian~\cite{Kosior2017_localization}
\begin{equation}
H = -\sum_{\langle i,j \rangle} J_{ij} \left( c_i^\dagger c_j + \text{H.c.} \right),
\label{eq:tb_fractal}
\end{equation}
where $c_i^\dagger$ ($c_i$) creates (annihilates) a fermion at site $i$. The hopping amplitude is nonzero ($J_{ij} = J$) only when sites $i$ and $j$ are connected in the RFL, including the additional links introduced through $p$. Previous work showed that such models provide a useful platform for studying localization-delocalization transitions in non-Euclidean geometries, where spectral dimension plays a central role~\cite{Kosior2017_localization} (see also a recent preprint~\cite{Li2026AndersonFractal} and other related works \cite{GarcaMata2020,Sierant2023, Manna2024, RojoFrancs2024, Bhattacharjee2025, Bhattacharya2026}). Here, by contrast, we show that the Hausdorff dimension primarily controls the ground-state entanglement scaling, but both dimensions are important for the dynamical generation of entanglement. Throughout this paper, we work in the dimensionless units where the energy and time scales are set by the tunneling amplitude $J$ and $\hbar / J$, respectively. 

\begin{figure}[t!]
  \centering
  \includegraphics[width= 0.49\textwidth]{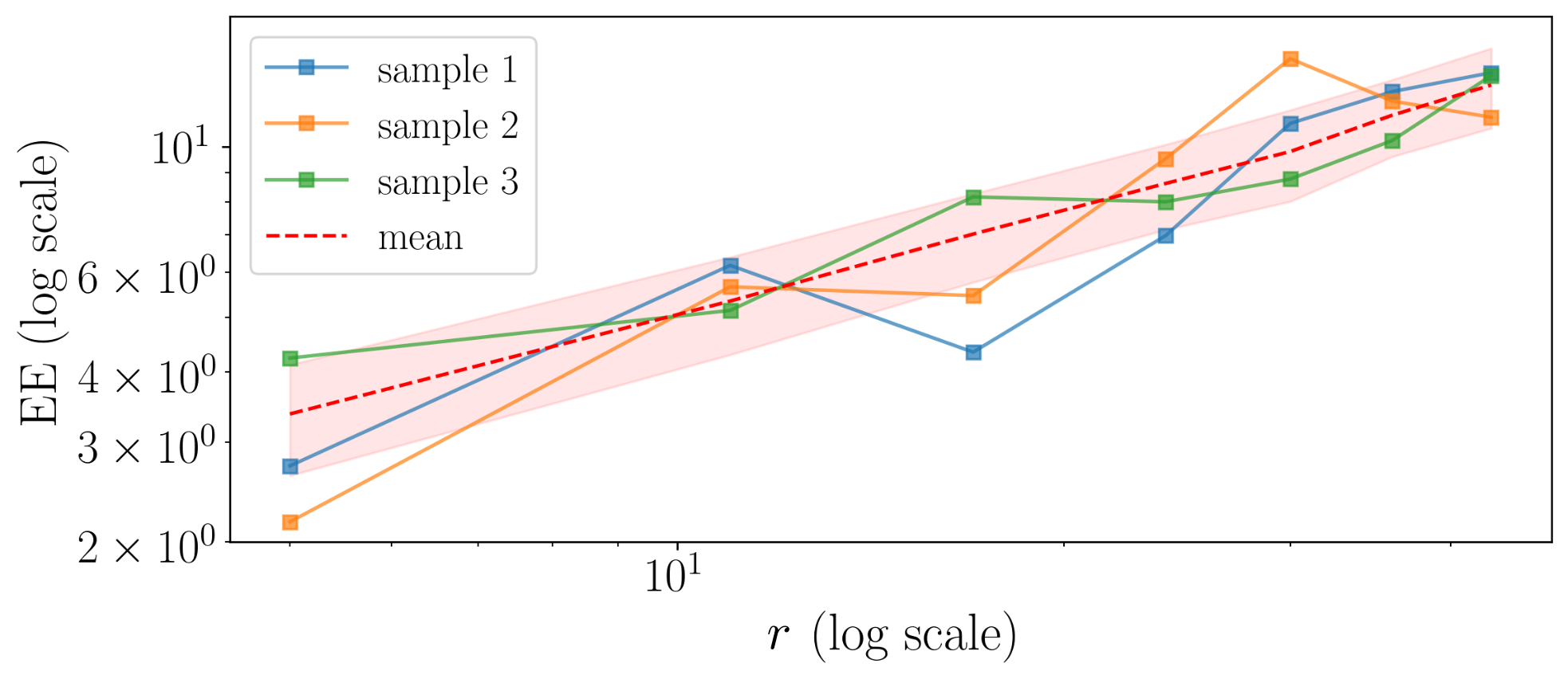}
  \includegraphics[width= 0.49\textwidth]{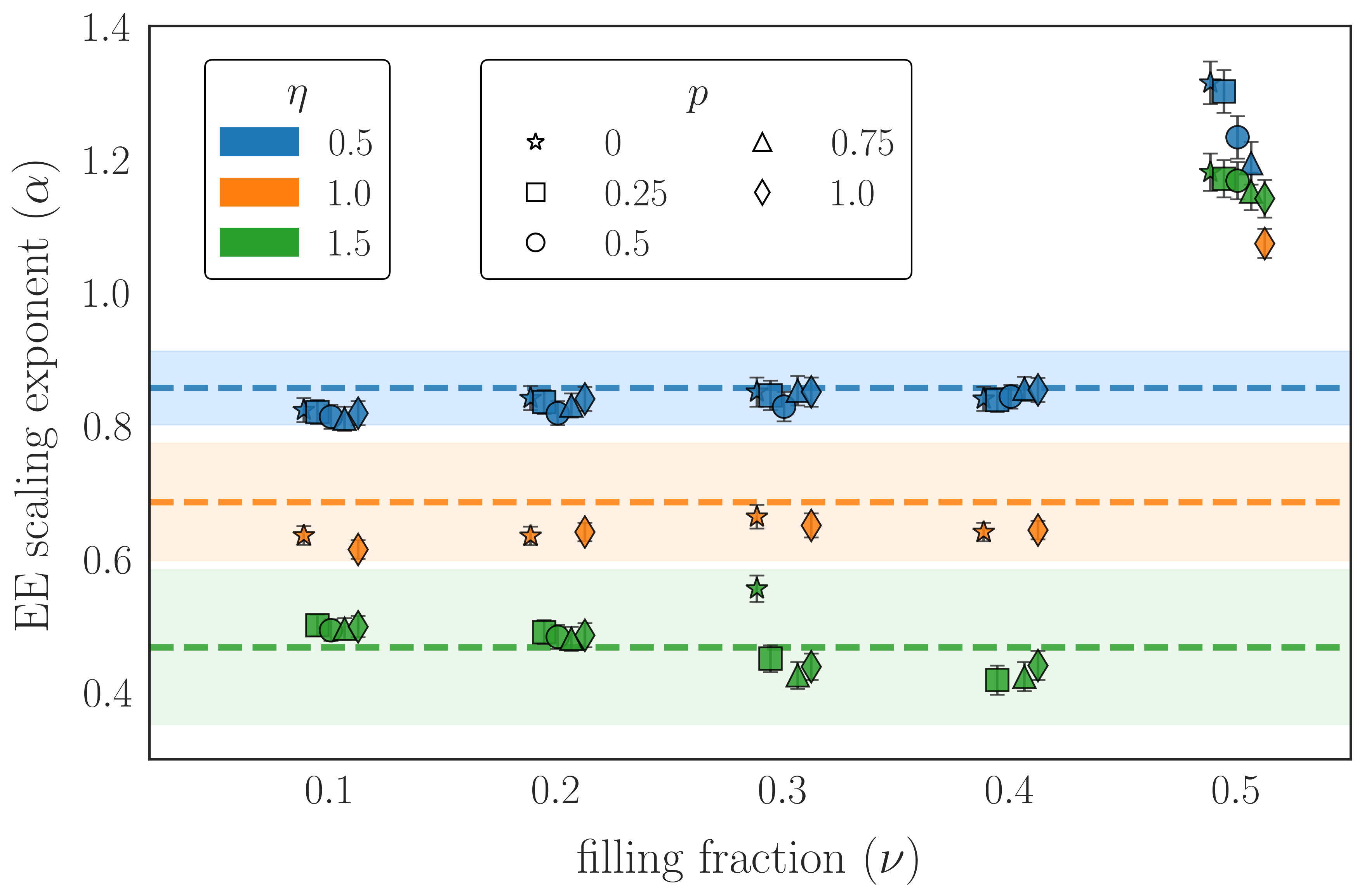}
  \caption{(top panel) Although the entanglement entropy (EE) shows strong sample-to-sample variability, its disorder average exhibits clear power-law behavior. (bottom panel) Scaling exponent for the EE as a function of filling $\nu$ for different $(\eta,p)$ pairs. For each parameter set, we fit $S$ versus subsystem radius $r$ on log--log axes and extract an effective power-law exponent; symbols show averages over 100 independent realizations. Dashed lines indicate area-law scaling $\alpha=d_\mathrm{H}-1$, and shaded regions denote one standard deviation across samples.}
  \label{fig3}
\end{figure}

\textit{Ground-state entanglement.---}
We characterize the bipartite entanglement entropy (EE) of a subsystem $A_r$, defined as the the set of sites $i$ within radius $r$ around a chosen origin $i_0$,
\begin{equation}
A_r = \{ i : ||i-i_0|| \le r \},
\end{equation}
where $|| \cdot ||$ denotes Eucliedean norm. For a many-body ground state $|\Psi_0\rangle$ at filling $\nu$, the EE is given by the von Neumann entropy
\begin{equation}
  S(r) = -\text{Tr} \rho_{A_r} \ln \rho_{A_r}, \quad \rho_{A_r} = \text{Tr}_{\bar{A}_r} |\Psi_0\rangle \langle \Psi_0|,
\end{equation}
where $\rho_{A_r}$ is the reduced density matrix and $\bar{A}_r$ denotes the complement of $A_r$.
For noninteracting fermions, $S(r)$ is efficiently evaluated via the restricted one-body correlation matrix $C_{ij} = \langle \Psi_0 | c_i^{\dagger} c_j | \Psi_0 \rangle$ for $i,j \in A_r$ \cite{Peschel2003}. Given the eigenvalues $\{\lambda_m\}$ of the correlation matrix, the entropy reads
\begin{equation}
S(r) = -\sum_m \left[ \lambda_m \ln \lambda_m + (1-\lambda_m) \ln (1-\lambda_m) \right].
\end{equation}
We use this procedure to analyze the scaling of $S(r)$ with subsystem size $r$ across a range of fillings $\nu$. 

For each parameter set $\{\eta, p\}$, we average $S(r)$ over $10^2$ independent lattice realizations to suppress the strong sensitivity of the entanglement to microscopic configuration details. We find that the disorder-averaged entropy scales algebraically with subsystem size, $\langle S(r) \rangle \propto r^\alpha$, as shown in Fig.~\ref{fig3}, where results for different subsystem sizes $r$ collapse for the same $\{\eta, p\}$. Notably, we find no evidence of logarithmic enhancement, which we attribute to the lack of a conventional Fermi surface. Away from half-filling, the exponent is consistently $\alpha \approx d_\mathrm{H} - 1$, where $d_\mathrm{H}$ is the Hausdorff dimension. This generalized area law persists independently of the filling $\nu$ and connectivity $p$. Near half-filling, however, the EE exhibits anomalous scaling. Same behavior, which 
originates from the high spectral degeneracy at the band center \cite{Kosior2017_localization}, was found and extensively studied in percolating clusters \cite{Pachhal2024}. 
% , a feature whose precise nature remains a subject for future study.

\textit{Entanglement entropy dynamics.---}
To investigate the growth of entanglement under the fractal Hamiltonian in Eq.~\eqref{eq:tb_fractal}, we consider a global quench starting from an uncorrelated checkerboard product state:
\begin{equation}
  |\Psi(0)\rangle = \prod_{j\in \mathcal{C}} c_j^{\dagger} |0\rangle,
\end{equation}
where $\mathcal{C}$ denotes the occupied sites of a checkerboard sublattice on the square embedding restricted to the random fractal. In clean, short-range Euclidean systems, quench dynamics are typically governed by the quasiparticle picture, leading to an initial linear rise in entanglement for times $t<r/v_{\mathrm{max}}$, where $v_{\mathrm{max}}$ is the maximum quasiparticle velocity~\cite{Calabrese2005_quench,Lieb1972_bound,Alba2017_quasiparticle}. In contrast, transport on random fractal lattices can be fundamentally different.  

\begin{figure}[t!]
  \centering
  \includegraphics[width= 0.49\textwidth]{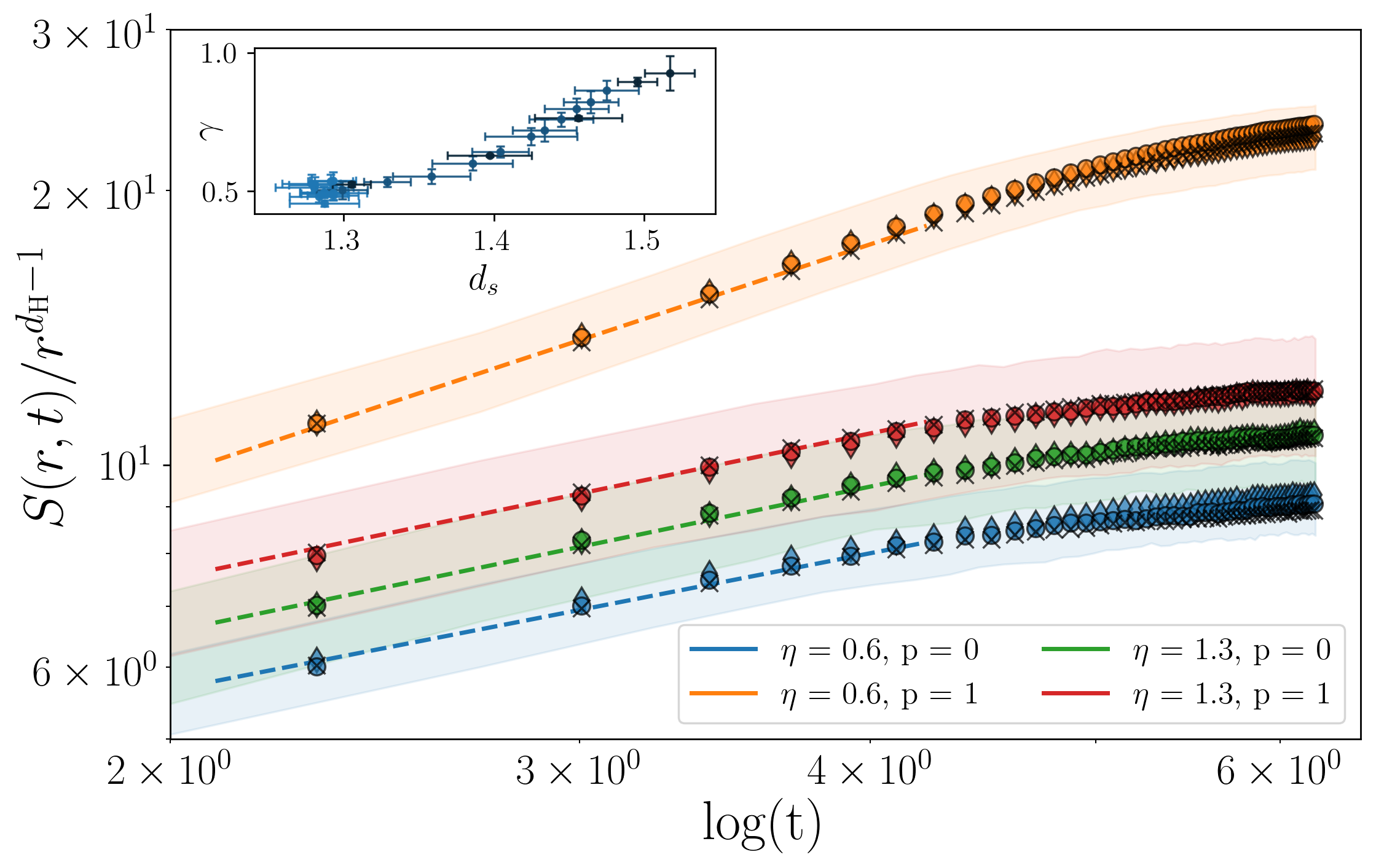}
  \caption{Entanglement-entropy growth following a global quench from the checkerboard product state on random fractal lattices. The entropy $S(r,t)$ is shown for subsystems of graph radius $r$ in the intermediate-time regime, after the microscopic short-time regime and before finite-size effects become important. The data are consistent with the scaling form $S(r,t)\propto a(r)s(t)$, with $a(r)=r^{d_{\mathrm H}-1}$ and a logarithmically slow temporal growth $s(t)\sim [\log(t)]^{\gamma}$. Symbols show disorder averages over 100 independent realizations for several subsystem radii $r$ close to half the system size. (inset) The exponent $\gamma$ as a function of the spectral dimension, obtained from numerical fitting. The color depth of the inset data points represents different values of $\eta \in \{0.5,0.6,1.3\}$.}
  \label{fig4}
\end{figure}

\begin{figure}[t!]
  \centering
  \includegraphics[width= 0.49\textwidth]{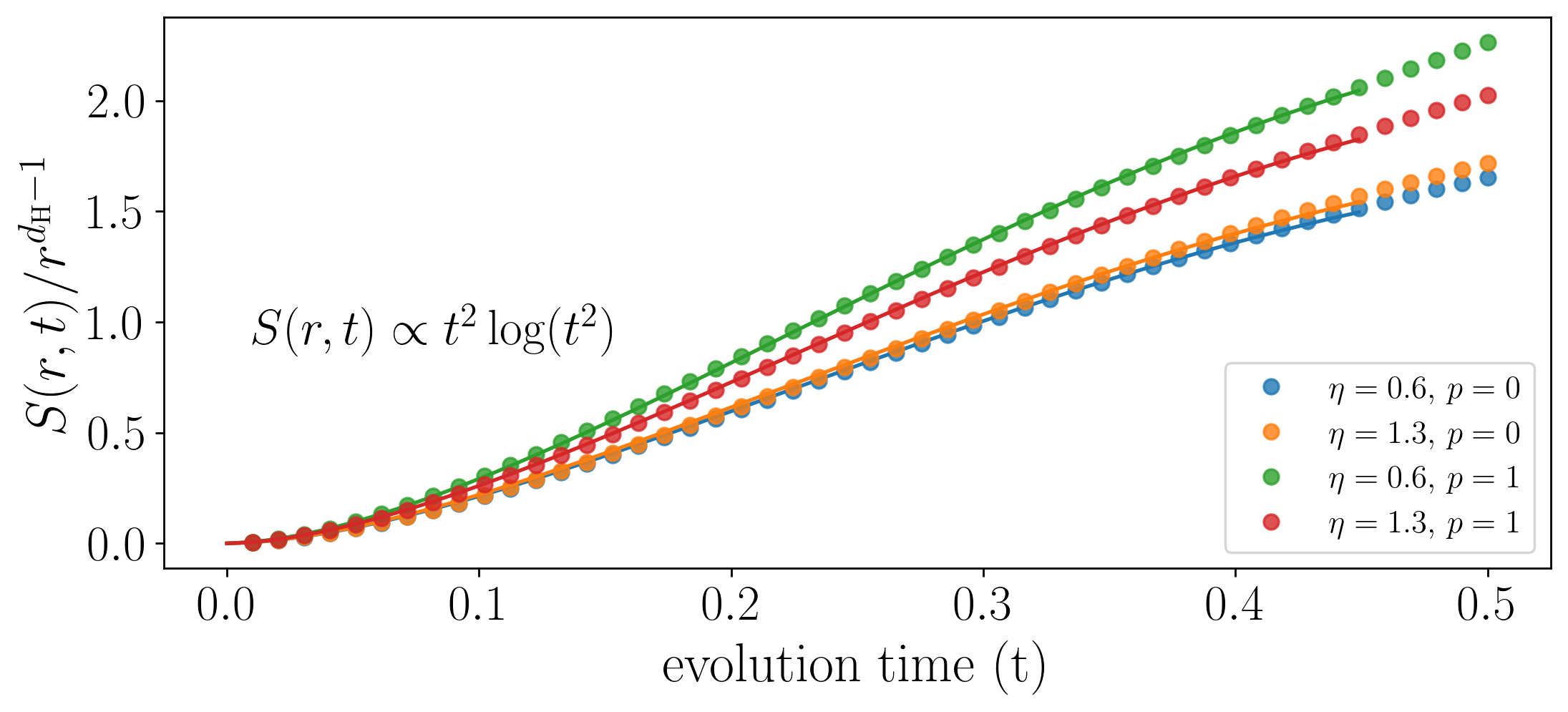}
  \caption{Short-time entanglement growth after the global quench from the checkerboard product state on random fractal lattices. The entropy $S(r,t)$ is shown in the early-time regime for representative subsystem radii and fractal parameters. At these times, the dynamics have not yet resolved the underlying fractal geometry, so the growth is governed primarily by local hopping processes and follows the expected universal short-time form $s(t)\sim t^2\log(t^2)$. This behavior is independent of both the Hausdorff and spectral dimensions within the numerical accuracy of our data. Symbols denote disorder-averaged results over independent lattice realizations, and solid curves show nonlinear fits to the short-time scaling form.}
  \label{fig5}
\end{figure}

We emphasize three main features of this section. First, up to the onset of finite-size saturation, the entanglement entropy exhibits the scaling form
\begin{equation}
  S(r,t) \propto a(r;d_{\mathrm H})\, s(t;d_{\mathrm s}),
\end{equation}
which may be viewed as a natural generalization of the corresponding free-scalar-field result in uniform space~\cite{Cotler2016GlobalQuench}. This expression reveals a separation between geometric and dynamical aspects of entanglement growth on fractal lattices. The Hausdorff dimension $d_{\mathrm H}$ controls the spatial scaling of entanglement, while the spectral dimension $d_{\mathrm s}$ governs the temporal evolution. Here, $a(r;d_{\mathrm H})=r^{d_{\mathrm H}-1}$ is the fractal area-law factor, and $s(t;d_{\mathrm s})$ is a universal scaling function that depends only on the spectral dimension. This universality is evidenced by the collapse of numerical results for different subsystem sizes $r$ close to half the system size, as indicated by the different symbols in Fig.~\ref{fig4}.

Second, in the intermediate-time regime, where $t$ exceeds the microscopic tunneling time but remains short enough to avoid finite-size effects, we do not observe a transient ballistic quasiparticle regime. Instead, the dynamics remain sub-ballistic, with
\begin{equation}
 s(t;d_{\mathrm s})\propto [\log(t)]^{\gamma(d_{\mathrm s})},
\end{equation}
where $\gamma$ is a dynamical exponent. As shown in Fig.~\ref{fig4}, the data for different pairs of fractal parameters $(\eta,p)$ consistently exhibit logarithmic growth in this regime. This behavior is reminiscent of disordered free-fermion systems~\cite{Sirker2019PRB}. Crucially, however, the exponent $\gamma$ depends nonlinearly only on the spectral dimension $d_{\mathrm s}$, and is essentially independent of the Hausdorff dimension.

Third, at very early times, $t\lesssim 0.5$, i.e., before a single tunneling time has elapsed, we expect $s(t;d_{\mathrm s})$ to be insensitive to microscopic details of the lattice because the dynamics have not yet resolved the underlying topology and interference effects remain negligible. In this regime, lattice systems are expected to exhibit the characteristic $t^2\log(t^2)$ growth of entanglement entropy associated with the hopping term mixing local Hilbert spaces. Our numerical data in Fig.~\ref{fig5} are consistent with this expectation.

These findings show that nonequilibrium entanglement growth serves as a sensitive dynamical probe of fractal topology, with the dependence of $\gamma$ on connectivity $p$ and filling $\eta$ providing a route to disentangling the interplay between spectral properties and quantum-information spreading.

%  because evolution time is shorter than
% a single tunneling time and the system does not yet resolve the underlying topology and interference effects are negligible. 

\textit{Summary and conclusions.---}
We studied the entanglement properties of free fermions on random fractal lattices generated by a dielectric-breakdown growth process. By varying the growth parameter and the probability of adding missing links, we accessed a broad family of geometries with tunable Hausdorff and spectral dimensions and used them to examine both ground-state entanglement and post-quench entanglement growth. For the ground state, we computed the bipartite entanglement entropy from the restricted correlation matrix and found that, over a wide parameter range, the entropy follows a robust power-law dependence on subsystem size. Away from half filling, the corresponding exponent is governed primarily by the Hausdorff dimension and is consistent with a generalized area law, without the logarithmic enhancement familiar from Euclidean free-fermion systems.
Near half filling, however, the scaling becomes anomalous: the effective exponent develops a more pronounced dependence on connectivity and spectral properties. This behavior points to a distinct entanglement regime associated with the special structure of states near the band center.
% whose microscopic origin deserves further investigation.

In the dynamical setting, we further found that a global quench from an uncorrelated checkerboard state leads to logarithmically slow entanglement growth over an extended intermediate-time window. The dynamics are well described by a factorized scaling form in which the subsystem-size dependence is set by a fractal area factor controlled by $d_{\mathrm H}$, whereas the temporal growth is governed primarily by the spectral dimension $d_{\mathrm s}$. This separation suggests that the geometry controls two distinct aspects of information spreading: the Hausdorff dimension fixes the amount of boundary through which correlations can be generated, while the spectral dimension determines how efficiently excitations explore the ramified graph. The resulting logarithmic growth is therefore not a consequence of interactions, onsite disorder, or many-body localization, but instead arises from the hierarchical connectivity and anomalous diffusion intrinsic to the random fractal lattice. 

More broadly, our results establish random fractal lattices as a promising arena for uncovering genuinely non-Euclidean entanglement phenomena, and they open a route toward systematically exploring how geometry alone can reshape quantum correlations, transport, and information spreading in complex many-body systems.

\textit{Acknowledgments.---}
We are grateful to Krzysztof Sacha for introducing us and making this collaboration possible. A.K. acknowledges support from the Polish National Agency for Academic Exchange (NAWA) under the Polish Returns Programme (Grant No. BPN/PPO/2024/1/00010/U/00001). J.W. acknowledges support from the Australian Research Council through a Future Fellowship (Grant No. FT230100229). 

\textit{Data availability.---}
The data presented in this article are available from \cite{zenodo}.

\bibliographystyle{apsrev4-2}
% \bibliography{fractals_refs.bib}

\clearpage

\begin{widetext}
\begin{center}
  \textbf{END MATTER}
\vspace{0.5cm}
\end{center}
\end{widetext}

\textit{Numerical details of growth algorithm.---}
In the numerical implementation of the dielectric-breakdown growth algorithm, the auxiliary potential $\phi$ is defined on a square Euclidean grid of linear size $L$. Depending on the growth parameter $\eta$ and the typical spatial extent of the generated cluster, we use grids with $L$ between 500 and 1000 to minimize boundary effects. At each growth step, the discrete Laplace equation is solved iteratively on this grid with boundary conditions imposed by the existing cluster.

\begin{figure}[b]
  \centering
  \includegraphics[width=0.5\textwidth]{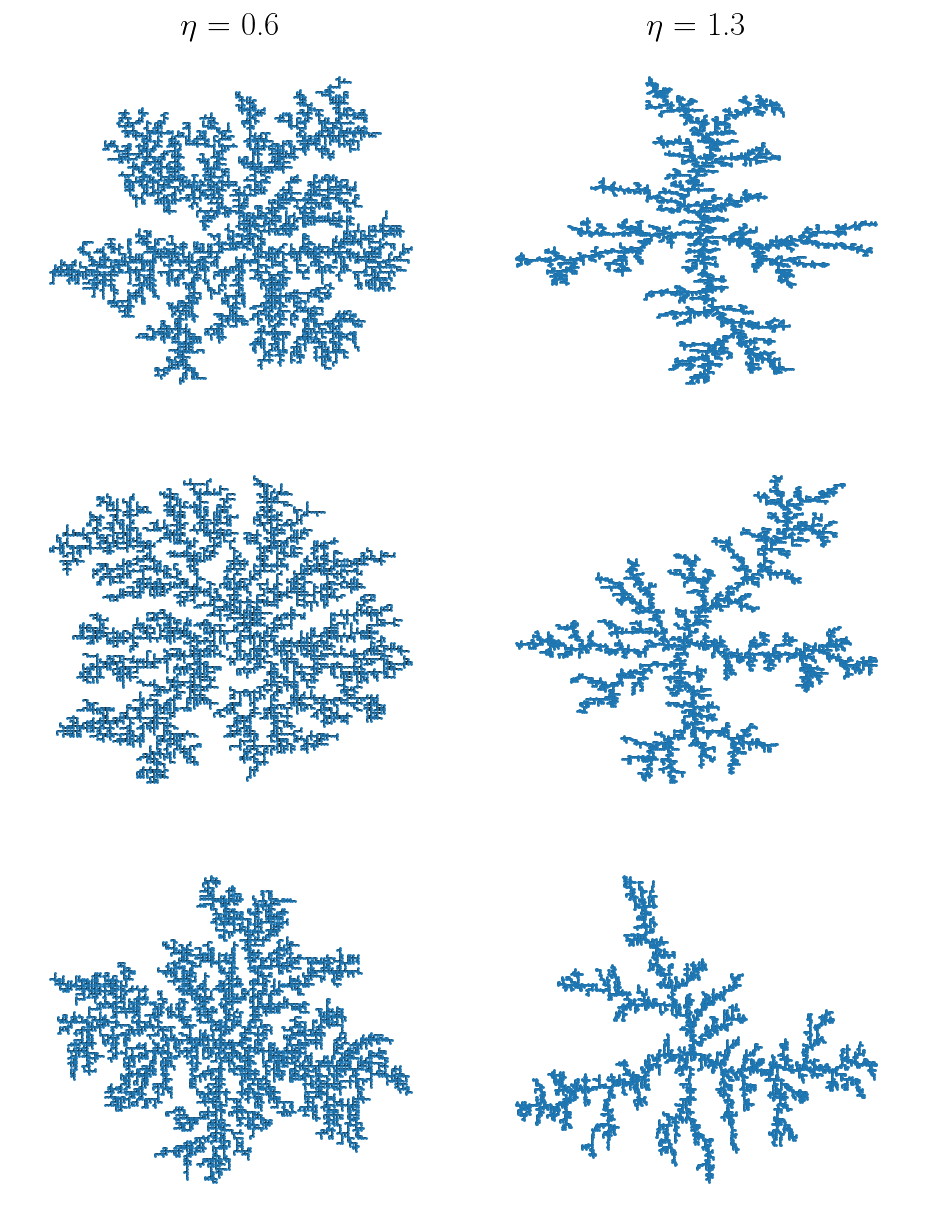}
  \caption{Representative random fractal lattice realizations generated by the dielectric-breakdown growth algorithm for two values of the growth parameter $\eta$. The left column shows samples for $\eta=0.6$, which produce denser clusters with larger Hausdorff dimension, while the right column shows samples for $\eta=1.3$, where growth is more strongly concentrated at the tips and the resulting clusters are more branched. Each row corresponds to an independent realization.}
  \label{fig:app_fractals}
\end{figure}

 Convergence of the relaxation procedure is monitored using the residual sum of squares (RSS), defined as 
\begin{equation}
  \mathrm{RSS}=
\sum_i[\phi_i^{(n+1)}-\phi_i^{(n)}]^2
\end{equation}
where $\phi_i^{(n)}$
is the potential at grid site $i$ after the $n$th relaxation iteration. The iteration is stopped once the RSS falls below the tolerance $\epsilon=10^{-5}$. After convergence, a new boundary site is selected according to the growth probability defined in the main text, added to the cluster, and the potential field is updated before the next growth step. Representative lattice realizations generated with this procedure are shown in Fig.~\ref{fig:app_fractals}.

\begin{figure}[hbt]
  \centering
  \includegraphics[width=0.5\textwidth]{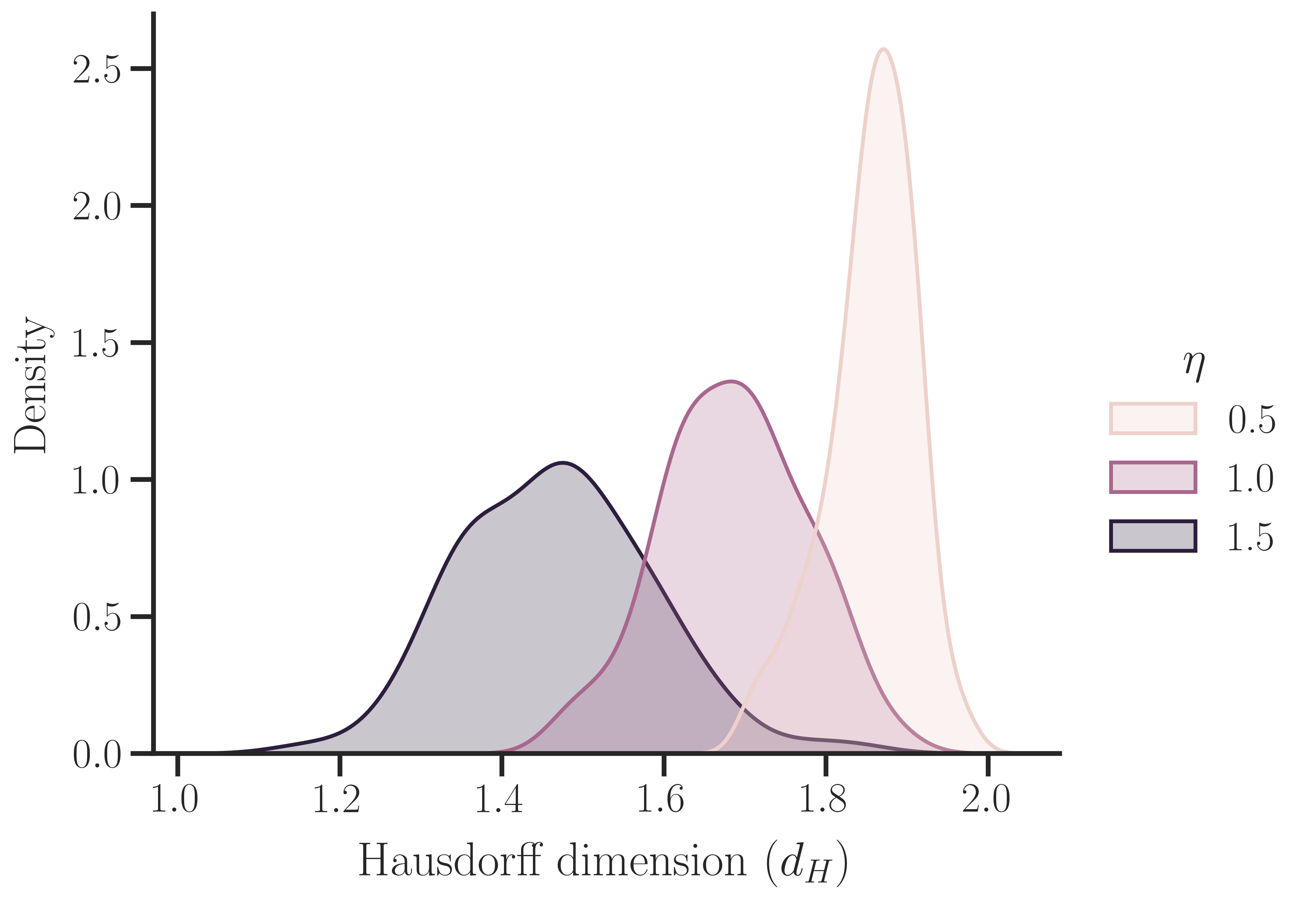}
  \caption{Distribution of the Hausdorff dimension $d_{\mathrm H}$ extracted from independent random fractal lattice realizations for different values of the growth parameter $\eta$. Increasing $\eta$ produces more branched, lower-dimensional clusters, shifting the distribution of $d_{\mathrm H}$ to smaller values. The widths of the distributions quantify sample-to-sample fluctuations within each ensemble.}
  \label{fig:app_dh_distribution}
\end{figure}

\begin{figure}[hbt]
  \centering
  \includegraphics[width=0.5\textwidth]{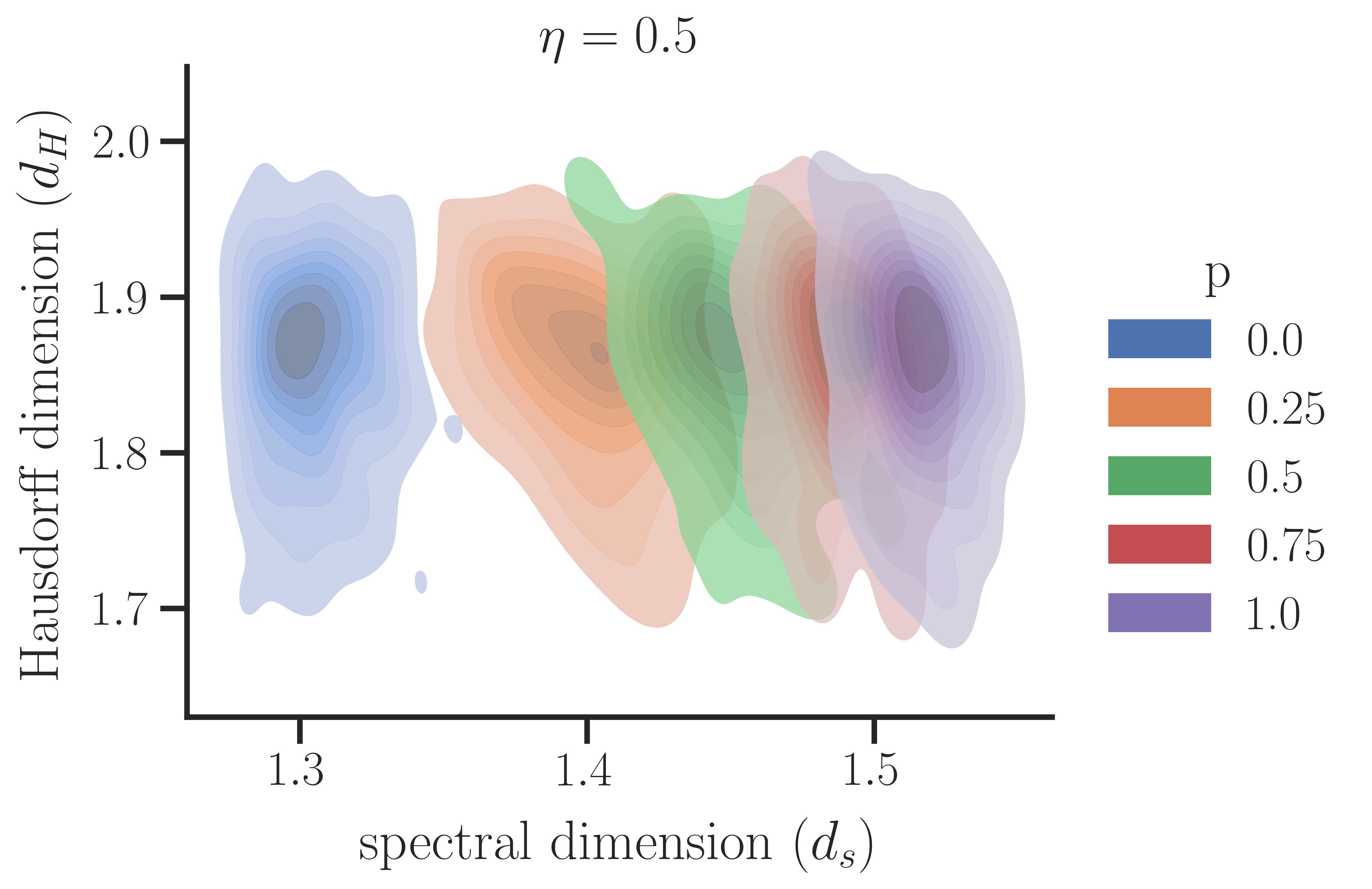}
  \includegraphics[width=0.5\textwidth]{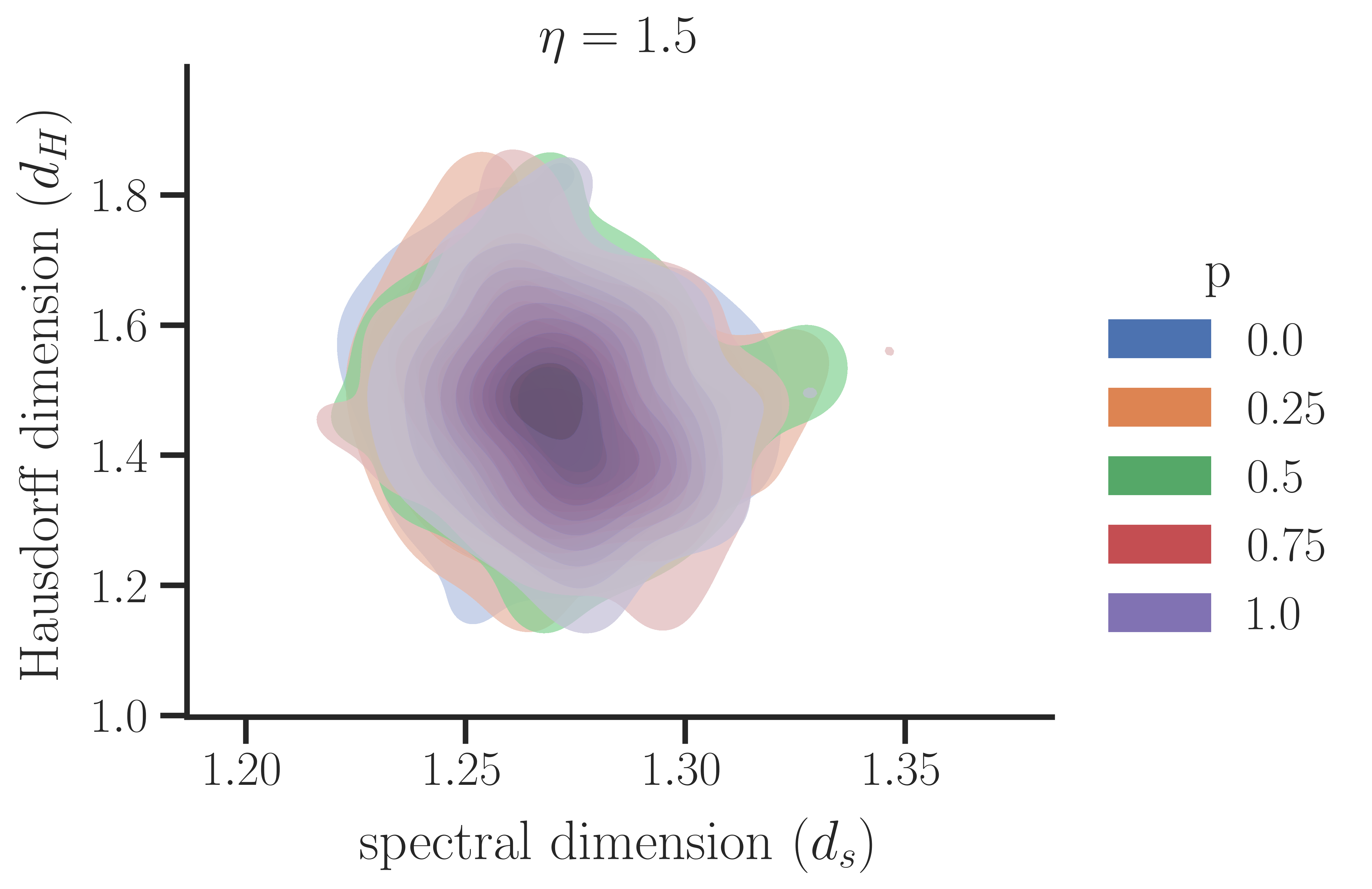}
  \caption{Distribution of the fractal dimensions extracted from independent random fractal lattice realizations. The Hausdorff dimension $d_{\mathrm H}$ is obtained from the scaling of the graph volume $V(r)\sim r^{d_{\mathrm H}}$, while the spectral dimension $d_{\mathrm s}$ is extracted from the random-walk scaling $\Omega(t)\sim t^{d_{\mathrm s}/2}$. The histograms quantify sample-to-sample fluctuations for fixed growth and connectivity parameters.}
  \label{fig:app_fractal_dimensions}
\end{figure}

\begin{figure}[hbt]
  \centering
  \vspace{1cm}
  \includegraphics[width=0.5\textwidth]{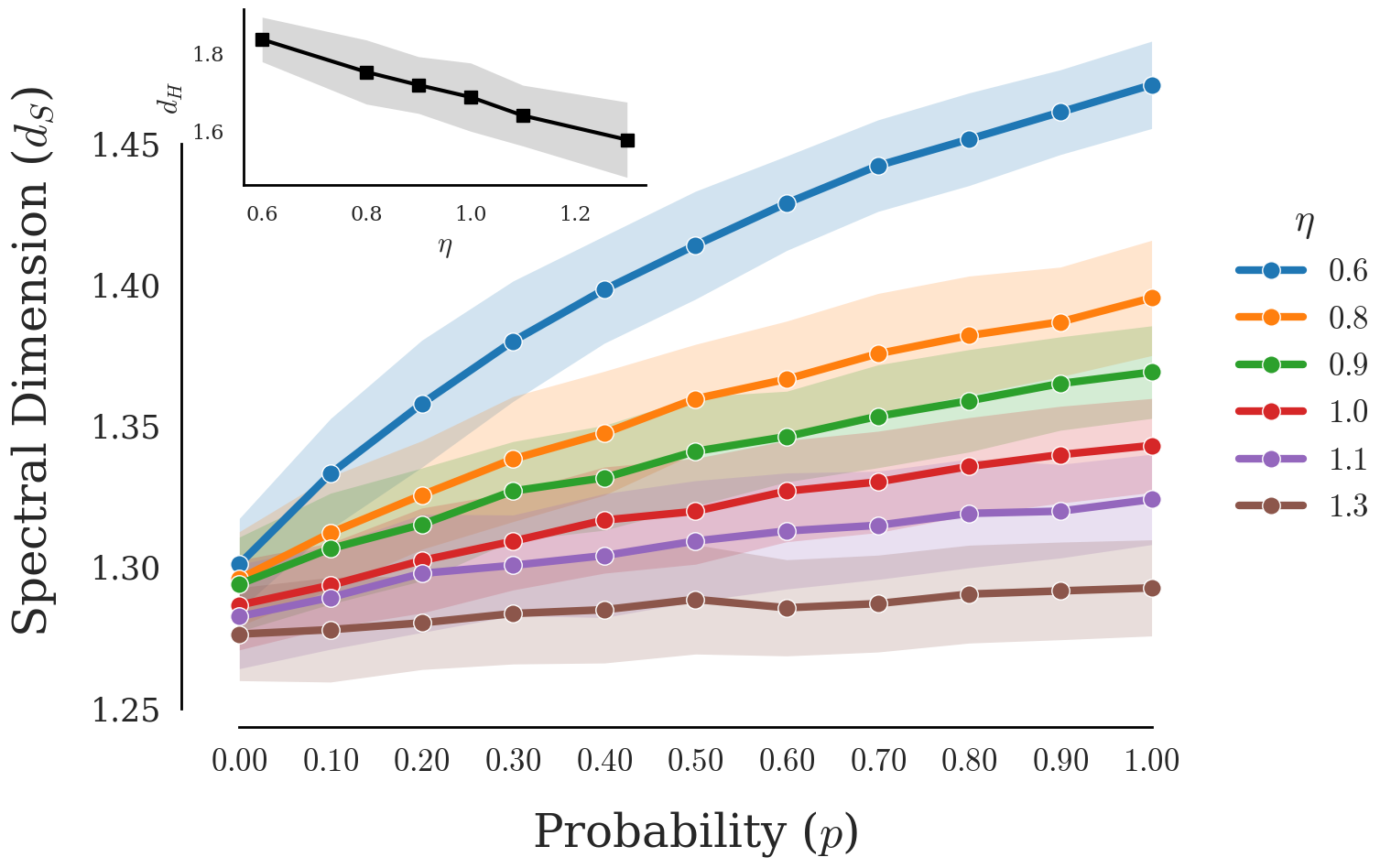}
  \caption{Extracted spectral dimensions as a function of the connectivity parameter $p$ for different values of the growth parameter $\eta$. Symbols show averages over realizations, and shaded regions denote one standard deviation. Inset: extracted spectral dimension as a function of the growth parameter $\eta$.}
  \label{fig2}
\end{figure}

\begin{table}[hbt]
\caption{Parameter sets and extracted fractal dimensions used in the numerical simulations. For each growth parameter $\eta$, the added-link probability $p$ is sampled uniformly from 0 to 1 in steps of 0.1. The Hausdorff dimension is reported as the ensemble mean and standard deviation, while the spectral-dimension column gives the range of mean values obtained as $p$ is varied.}
\label{tab:app_parameters}
\begin{ruledtabular}
\begin{tabular}{cccc}
$\eta$ & $p$ range & $d_{\mathrm H}$ & $d_{\mathrm s}$ range \\
\hline
0.6 & 0.0--1.0 & $1.835\pm 0.057$ & 1.301--1.471 \\
0.8 & 0.0--1.0 & $1.752\pm 0.082$ & 1.296--1.395 \\
0.9 & 0.0--1.0 & $1.719\pm 0.072$ & 1.294--1.369 \\
1.0 & 0.0--1.0 & $1.688\pm 0.087$ & 1.287--1.343 \\
1.1 & 0.0--1.0 & $1.641\pm 0.078$ & 1.283--1.324 \\
1.3 & 0.0--1.0 & $1.579\pm 0.096$ & 1.276--1.293 \\
\end{tabular}
\end{ruledtabular}
\end{table}

\textit{Fractal-dimension distributions.---}
Because the growth process is stochastic, individual lattice realizations generated with the same parameters $\eta$ and $p$ differ at the microscopic level, as illustrated in Fig.~\ref{fig:app_dh_distribution}. We therefore extract the fractal dimensions separately for each realization and then average over the ensemble. The Hausdorff dimension is controlled primarily by the growth parameter $\eta$: larger $\eta$ favors tip growth and produces more tenuous, branched clusters with smaller Hausdorff dimension $d_{\mathrm H}$. The resulting histograms are approximately normally distributed, with widths that decrease as $\eta$ increases.

The spectral dimension can be tuned further by the added-link probability $p$, which modifies the effective connectivity of the graph and hence its diffusion properties. The corresponding distributions are shown in Fig.~\ref{fig:app_fractal_dimensions}. For larger $\eta$, the clusters are sparser and contain fewer missing nearest-neighbor links that can be added; consequently, the accessible variation of $d_{\mathrm s}$ with $p$ becomes smaller. These distributions provide the statistical input for the disorder-averaged entanglement results discussed in the main text. The parameter sets and extracted dimensions are summarized in Table~\ref{tab:app_parameters} and plotted in Fig.~\ref{fig2}.

\end{document}